# A method of real-time analysis for stray light uniformity of optical telescope

Taoran Li[a], Jianfeng Tian[a], Xue Yang[b,c], Ying Wu[a,c]

[a]Key Laboratory of Optical Astronomy, National Astronomical Observatories, Chinese Academy of Sciences, 100101, Beijing, China;
[b]CAS Key Laboratory of Space Astronomy and Technology, National Astronomical Observatories, Chinese Academy of Sciences, 100101, Beijing, China;
[c]School of Astronomy and Space Science, University of Chinese Academy of Sciences, 100049, Beijing, China

## ABSTRACT

The stray light uniformity is one of the important factors affecting the signal-to-noise ratio of the optical astronomical telescope. It will cause regional differences in the background intensity of the detector image, resulting in a decrease of the differential photometry accuracy. The source that affects stray light uniformity is the inconsistency of the brightness of the sky background, which comes from moonlight, bright star, and city lighting pollution. During CCD reduction, the effect of background uniformity cannot be eliminated by dividing the flat field.

Star deletion method is used in real-time stray light analysis. It's very convenient to achieve a 'clear' background image without stars in MATLAB. A contour map of stray light distribution for each object image will be given to demonstrate the background uniformity directly. The stray light uniformity analysis method is implemented by the following steps: 1) CCD reduction, including preprocessing of an object image with bias and flat field; 2) Histogram generation, performing star subtraction automatically based on ADU value and frequency; 3) Background stray light contour map generation, stray light uniformity and other parameters calculations. This method will calculate the uniformity of image surface in real time, provide background intensity distribution, statistical data of the CCD image and suggestion on compare star selection during CCD data processing and improve the photometry accuracy.

## 1. INTRODUCTION

The large diameter and high precision orient the development of modern optical astronomical telescopes. The photometry accuracy of the observation data usually determines the credibility of scientific results. The differential photometry, which is the simplest of calibrations, allows simultaneously measuring the flux of a target and nearby stars in the same starfield[1] or relative photometry by comparing the brightness of the target to stars with given fixed magnitudes[2]. When using CCD photometry, both the target and comparison stars are observed at the same time with the same filters, because of which the atmosphere effect could be neglected. For time series observations, differential photometry is very convenient when plotting magnitude change over time of a variable star and is usually compiled into a light curve.

The stray light uniformity is one of the important factors affecting the signal-to-noise of optical astronomical telescope. It will cause regional differences of background intensity on focal plane, resulting in a decrease in the accuracy of differential photometry. This inhomogeneity mainly comes from the inconsistency sky background brightness caused by moonlight, bright star and urban lighting pollution.

The real-time processing of photometric data can improve observation effective. It is one of the future developments of astronomical data analysis. Previous data processing methods did not consider the accuracy variation by inhomogeneity sky background, which cannot be eliminated by flat field correction. Otherwise, the stray light distribution cannot be effectively detected by adjusting the contrast. Real-time detection of stray light uniformity can assist in the selection of reference stars in differential photometry and reduce the influence of background difference on accuracy.

This paper provides a real-time analysis method for stray light uniformity of optical telescope, which can calculate the uniformity on focal plane of the telescope in real time, show background intensity distribution and improve the photometry accuracy. Figure 1 shows the flowchart of analysis method.

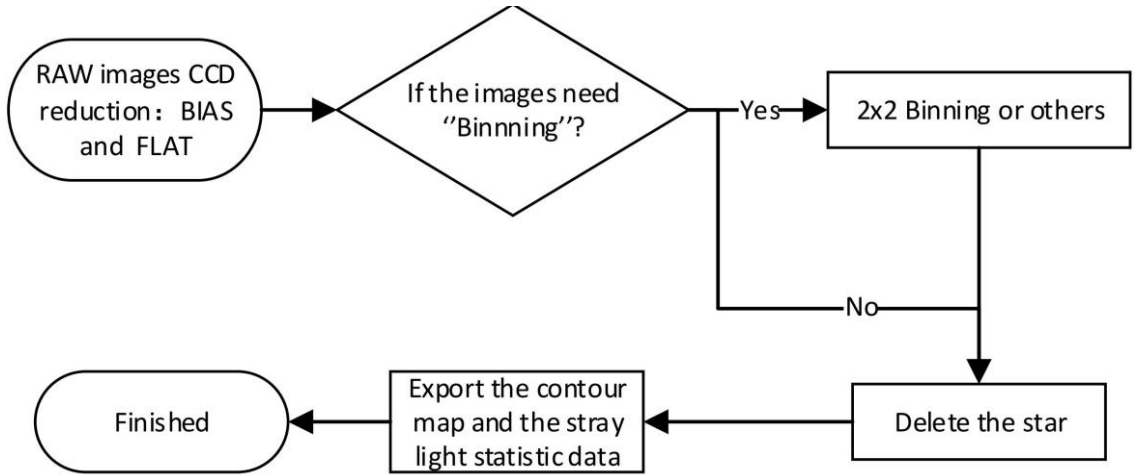

Figure 1 Flowchart of stray light uniformity analysis method

## 2. REAL-TIME STRAY LIGHT UNIFORMITY ANALYSIS

Normally, a set of auxiliary images will be taken before targets observations in order to run CCD reduction. These pre-observed images including Bias and Flat field could make the photometry of the object images accurate. The Bias is a frame of the shortest possible exposure, taken with the shutter closed. It represents the minimum noise in the CCD detector and camera circuitry[3]. The Flat field is an image of evenly illuminated field with the shutter open. It could correct different response causing by the inconsistent of the quantum efficiency and sensitivity of each pixel, and the optical noise in the system, such as dust motes on glass surfaces.

The raw object and auxiliary images are saved in local folder. The real-time analysis will identify different types of images by scanning the file names for real-time processing. The image types contain Bias, Flat field, standard star image and object image. When new standard star and object images are taken, the analysis system will detect it and perform CCD reduction automatically.

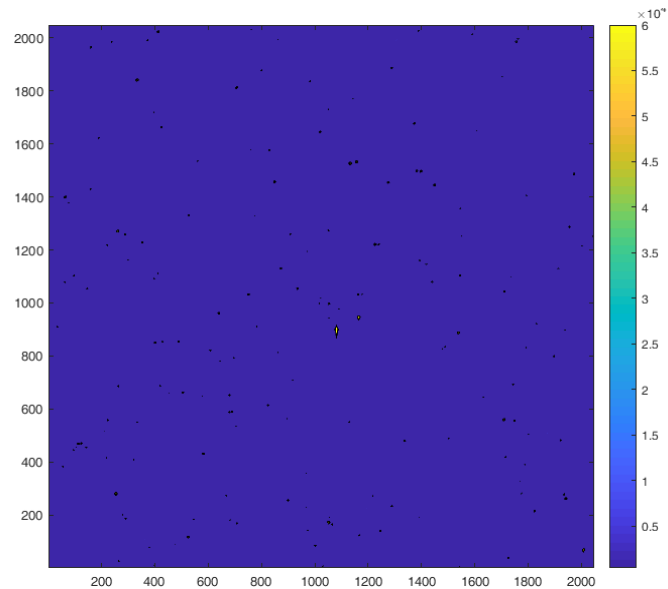

Figure 2 The contour map without any processes. The background information is submerged by bright stars (blue).

The contour map allows to show the distribution of stray light clearly and assist reference stars selection. The brightness of target stars and the reference stars are much higher than background due to the signal-to-noise ratio (SNR) demand of scientific objective. Making a contour map without any processes will give a terribly bright picture because the background stray light is submerged by the bright stars, resulting in indistinguishable stray light uniformity (figure 2). It may affect the stray light uniformity analysis. Meanwhile, the background uniformity of stray light detection is independent with the stars. Therefore, the stars are deleted to avoid producing errors in the uniformity analysis. The star deletion method is implemented by using histogram in MATLAB software. Approximation is adopted in star deletion that only minimum ADU (Analogue-to- Digital Unit)[4] value of the star is found without extracting every star.

## 2.1 Star deletion

Star deletion will deal with the images that experience CCD reduction. The first step is fits image reading. MATLAB reads the fits file as a two-dimensional matrix of the number of row pixels multiplied by the number of column pixels. The values in the matrix correspond to the ADU numbers of each pixel.

The second step is histogram making. The two-dimensional matrix must be transformed into a one-dimensional vector to calculate the ADU distribution by making a histogram. The histogram is plotted with the ADU numbers (abscissa, 0 to maximum ADU numbers of image) and its frequency (ordinate), at intervals of 100 ADUs, as shown in figure 3.

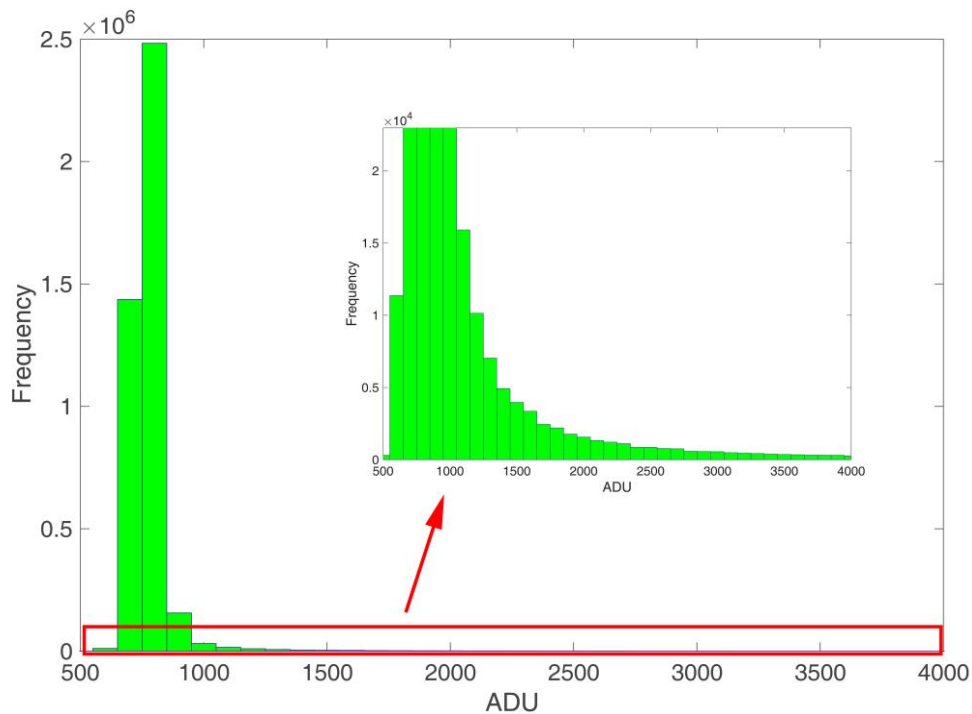

Figure 3 Histogram of an image by MATLAB, zoom in x-axis, the minimum and median ADUs are 433 and 768, respectively. The image in image (pointed by red arrow) is enlarged with y-axis.

Determining the thresholds of ADU value of stars and background is the third and most important step. The background pixels occupy a large percentage of the entire image. Also, the ADU value of the stars are always larger than that of background. The ADU value thresholds is automatically determined based on ADU frequency. The percentages of each interval of the histogram are calculating and each pixel in the interval with the percentage less than 1% is regarded as a star. If the telescope is pointed to a concentrated field like star cluster, the value of 1% can be increased. The percentage of each interval is accumulated from minimum ADU. When the accumulated value is greater than 99%, the position of that interval is recorded. The lower ADU limit of this interval is used as the thresholds of stars and background (maximum value of background or the minimum value of star).

Finally, all pixels which ADU value larger than the thresholds are replaced by the lower ADU limit of the interval (As table 1 shown, star = 1400, pixel(pixel>star) = star). The star deletion is successful.

Table 1 percentage of each interval of figure 2

| ADU range | Frequency | Percentage | Percentage(Accumulated) |
|---|---|---|---|
| 400-500 | 307 | 0.01% | 0.01% |
| 500-600 | 11360 | 0.27% | 0.28% |
| 600-700 | 1436541 | 34.25% | 34.53% |
| 700-800 | 2483256 | 59.21% | 93.73% |
| 800-900 | 156199 | 3.72% | 97.46% |
| 900-1000 | 31334 | 0.75% | 98.20% |
| 1000-1100 | 15906 | 0.38% | 98.58% |
| 1100-1200 | 10138 | 0.24% | 98.83% |
| 1200-1300 | 7034 | 0.17% | 98.99% |
| 1300-1400 | 4925 | 0.12% | 99.11% |
| 1400-1500 | 3969 | 0.09% | 99.21% |

## 2.2 Contour map making

Making a contour map of the background stray light is meant to calculate stray light uniformity. Figure 4 shows the contour map after star deletion procedure, with the star marked in yellow.

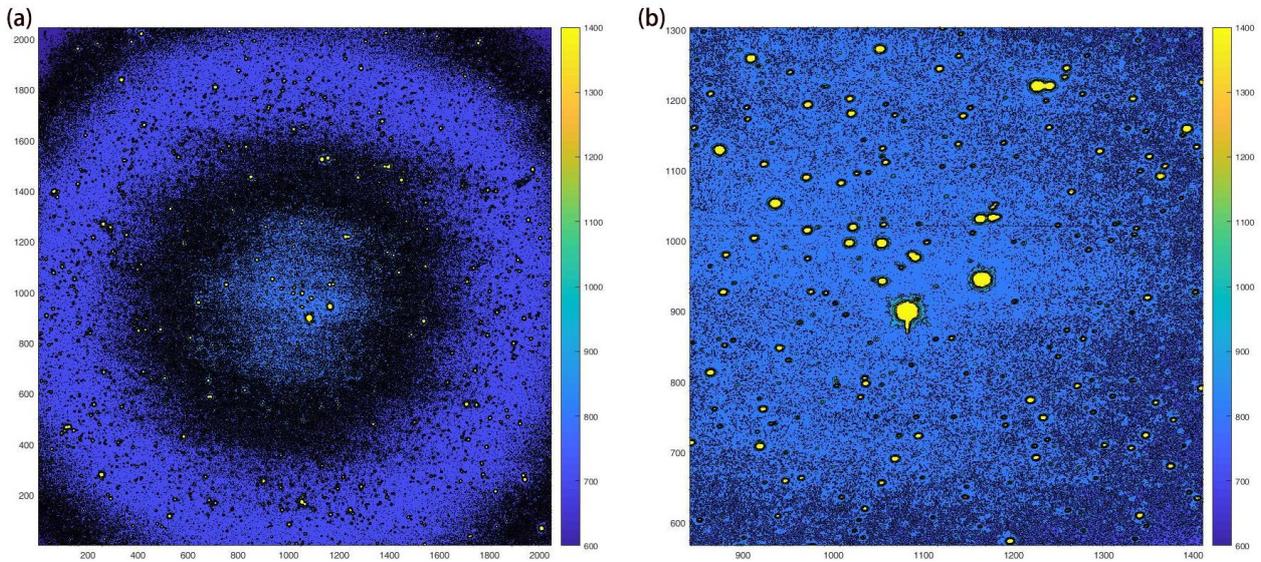

Figure 4 The contour map after star deletion. The star is marked in yellow (ADU=1400). (a): original scale; (b): enlarge the central portion.

The color bar in MATLAB is calculated from the maximum and minimum data. The minimum and maximum values are represented by the first and last colors in a colormap (such as Parula, Jet or HSV), respectively. It performs a linear transformation of the intermediate ADU values so that the image can be displayed in contour map within current data range[5]. The boundary of each area (color) is divided by solid black line. If the isolines are much dense in contour map, it seems to have more black areas as shown in figure 4(a). The black solid line can be seen clearly by enlarging the central portion of figure 4(a) (shown in figure 4(b)).

Figure 5 give two methods of stray light data statistics, dividing the image into four quadrants or rectangles with the origin coordinate of center point. The sequence of four quadrants is the same as Rectangular Plane Coordinate System. The sizes of four rectangles are 512×512, 1024×1024, 1536×1536, 2048×2048, respectively. Stray light information, such as stray light deviation (represented in background standard deviation), average of background, flux percentage and total flux, are collected in each quadrant or rectangular region. The standard deviation could quantify the amount of variation or dispersion of the background intensity[6], so the standard deviation of the background ADU values is used to evaluate the stray light uniformity.

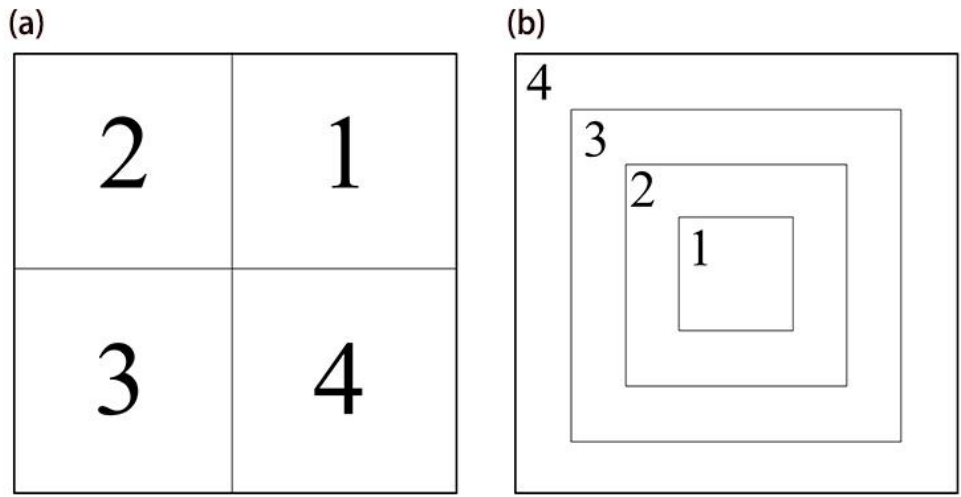

Figure 5 Sketches of four quadrants divided images (a) and multiple rectangles divided images (b).

Table 2 Statistics data of stray light in 4 quadrants

| Binning：No | Quadrant 1 | Quadrant 2 | Quadrant 3 | Quadrant 4 |
| --- | --- | --- | --- | --- |
| Uniformity (Standard deviation) | 87.9 | 84.9 | 86.7 | 89.1 |
| Max (Background) | 1400.0 | 1400.0. | 1400.0 | 1400.0 |
| Average | 782.8 | 784.5 | 773.9 | 776.3 |
| Min | 604 | 433 | 469 | 585 |
| Total flux | $8.2\times10^8$ | $8.2\times10^8$ | $8.1\times10^8$ | $8.1\times10^8$ |
| Percentage (%) | 25.1 | 25.2 | 24.8 | 24.9 |

Table 3 Statistics data of stray light in 4 rectangles

| Binning：No | Rectangle 1 | Rectangle 2 | Rectangle 3 | Rectangle 4 |
| --- | --- | --- | --- | --- |
| Uniformity (Standard deviation) | 86.5 | 81.9 | 81.0 | 87.3 |
| Max (Background) | 1400.0 | 1400.0 | 1400.0 | 1400.0 |
| Average | 842.4 | 823.1 | 801.8 | 779.4 |
| Min | 717.0 | 678.0 | 433.0 | 433.0 |
| Total flux | $2.2\times10^8$ | $8.6\times10^8$ | $1.9\times10^9$ | $3.3\times10^9$ |
| Percentage (%) | 6.8 | 26.5 | 57.9 | 100.0 |

The real-time analysis speed is correlated to the size of CCD sensor and the Computer performance. A large CCD sensor may need more time to make the contour map, with a risk of computer overload and software crash. The CCD binning is an image readout pattern that combines adjacent pixels and exports one pixel[7], offering benefits in computer processing speed while not affecting the stray light distribution. Even CCD binning will reduce spatial resolution, we may use this method to accelerate the real-time processing (Normally, when the CCD size larger than 4K×4K).

The final output results are contour map that clearly reveals the stray light distribution and table that shows the stray light data of each quadrant and rectangular region. Table 2 and Table 3 are some of the statistics data from figure 4(a).

## 3. CONCLUSIONS

The method described in this paper analyzes the uniformity of stray light in real time, providing the background distribution and statistical data on CCD image surface. It could improve the efficiency of reference star selection in the differential photometry.

The contour map obtained by this method allows to observe the stray light distribution directly. During photometry procedure, astronomers could choose the reference stars which locate in the region with the same color as the target star, avoiding background difference influence caused by stray light and improving the differential photometry accuracy. The real-time analysis method for stray light uniformity is proved to be feasible.

## ACKNOWLEDGEMENTS

This research was supported by grants from Joint Research Fund in Astronomy National Natural Science Foundation of China (No. U1831209) and National Natural Science Foundation of China, NSFC (Grant No. 11703043).